\documentclass[preprint,showpacs,preprintnumbers,amsmath,amssymb,nofootinbib]{revtex4}

\usepackage{graphicx}
\usepackage{dcolumn}
\usepackage{bm}

\def\beq{\begin{equation}}
\def\eeq{\end{equation}}
\def\bea{\begin{eqnarray}}
\def\eea{\end{eqnarray}}

\newcommand{\gsim}{ \mathop{}_{\textstyle \sim}^{\textstyle >} }

\usepackage{color}

\begin{document}

\preprint{UCI-HEP-TR-2010-18}

\title{Gamma Ray Line Constraints on Effective Theories\\ of Dark Matter}

\author{Jessica Goodman$^a$}
\author{Masahiro Ibe$^a$}
\author{Arvind Rajaraman$^a$}
\author{William Shepherd$^a$}
\author{Tim M.P. Tait$^a$}
\author{Hai-Bo Yu$^{a,b}$}

\affiliation{$^a$~Department of Physics \& Astronomy, University of California, Irvine, CA 92697}

\affiliation{$^b$~Michigan Center for Theoretical Physics, Department of Physics,\\
University of Michigan, Ann Arbor, MI 48109}

\date{August 31, 2010}

\begin{abstract}
A monochromatic gamma ray line results when dark matter particles in the galactic halo
annihilate to produce a two body final state which includes a photon.  Such a signal is
very distinctive from astrophysical backgrounds, and thus represents an incisive probe
of theories of dark matter.
We compare the recent null results of searches for gamma ray lines in the galactic center and
other regions of the sky with the predictions of effective theories describing
the interactions of dark matter particles with the Standard Model.
We find that the null results of these searches provide constraints on the nature of dark matter
interactions with ordinary matter which are complementary to constraints from other observables,
and stronger than collider constraints in some cases.
\end{abstract}

\pacs{95.35.+d, 14.80.-j}

\maketitle

\section{Introduction}

The nature and identity of dark matter (DM) is one of the most significant outstanding problems in
particle physics today. While cosmological measurements provide very compelling
evidence for the existence of dark matter, basic details such as
the mass of the dark matter particles and the nature of their interactions with
ordinary matter have remained elusive. Nevertheless, it is widely believed that dark matter
is a weakly interacting massive particle (WIMP).
WIMPs are highly motivated by the many models of physics beyond the Standard Model (SM)
which naturally produce stable candidates with the right properties and by the fact that
the relic density of dark matter can be explained as a natural consequence of the thermal
history of the Universe in such a scenario.

On the experimental side, if the dark matter has significant interactions with the Standard Model
particles, it may be detectable at direct detection experiments which search for signals of
the dark matter particle scattering off a nucleus.  In recent years, several
have reported anomalous results that are consistent with a light dark matter particle with
mass of order 10 GeV \cite{Bernabei:2010mq,Petriello:2008jj,Aalseth:2010vx}
(however, see \cite{CDMS-Si,Aprile:2010um,Sorensen:2010hq} as well),
leading to a surge of interest in the possibility of light dark matter \cite{Fitzpatrick:2010em,Kopp:2009qt,Kuflik:2010ah,Andreas:2010dz,Chang:2010yk,
Essig:2010ye,An:2010kc,Barger:2010yn,Hooper:2010uy}.
Another experimental probe which is
particularly effective for light dark matter is to look for signals of dark matter production at
colliders.  WIMP dark matter particles which couple to quarks and gluons can be pair
produced along with SM radiation from the SM initial state, leading to a signal involving a
high energy standard particle (such as a jet of hadrons or a photon) recoiling against
missing  momentum from the undetected
WIMPs \cite{Birkedal:2004xn,Cao:2009uw,Beltran:2010ww,Goodman:2010yf,Bai:2010hh}.
Colliders offer the advantage that the initial state and predicted
backgrounds are under exquisite control, which makes
the interpretations for given theories of WIMPs relatively straightforward.

A third broad class of experimental probes looks for the products of dark matter annihilating in
the halo of our galaxy.  The annihilation of WIMPs can produce distinctive
and/or high energy SM particles which may stick out over the expected background from
astrophysical sources.  Such processes have the advantage that they search for the actual
primordial dark matter particles in the cosmos, but this is also a disadvantage in that the rate
from a given part of the sky depends on the amount of dark matter (squared) along the
line of sight, which is not very well known.
An even more serious challenge to searches of this type is the fact that in
many cases the astrophysics is not very well understood, and thus the
expected background is difficult to quantify.

Distinguishing a signal of WIMP annihilation from astrophysics is much easier if there are
multiple handles defining the signal and/or features which even the uncertain
background has difficulty producing.  From both points of view, WIMP annihilations
producing monochromatic lines in the spectrum of high energy gamma rays are a promising
avenue 
\cite{Mack:2008wu,Bergstrom:1997fh,Bergstrom:2004nr,Dudas:2009uq,Perelstein:2006bq}.
Gamma rays have the advantage that they have a mean free path in the galaxy which is long
enough that they generally point back to their sources.  Thus, in searching for WIMP annihilations,
the origin and morphology of a given signal can be compared with expectations for the
distribution of dark matter and background sources.  In addition, high energy
photons are easy to detect, with the Fermi LAT satellite
rapidly improving our understanding of celestial gamma rays with energies ranging from
100's of MeV to TeV.

Dark matter annihilations into SM final states generically produce gamma rays, which
may be either radiated from charged final state particles or result from $\pi^0$ decays
originating
from the hadronization of colored final state particles.  These ``continuum" photons have
an energy spectrum which cuts off at the mass of the WIMP.  However, because the
photons are dominantly produced as secondary (or tertiary) particles in the annihilation,
the spectrum is highly peaked at low energies, and the presence of the cut-off may be
difficult to ascertain in the data.  A far more striking signature arises from
gamma ray lines, which result when WIMPs annihilate into a two particle final state
including a photon.  Denoting the WIMPs as $\chi$, the reaction
$\chi \chi \rightarrow \gamma X$ produces a photon with energy,
\bea
E_\gamma & = & m_{\chi} \left( 1 - \frac{M_X^2}{4 m_{\chi}^2} \right) ~.
\label{eq:Eline}
\eea
At large enough energies, it would be difficult to imagine such a line resulting from
any conventional astrophysical source.
Since WIMPs are forbidden by observations from having large couplings to the photon,
line emission processes are usually induced at loop level in perturbation theory.  Thus, their
rates are expected to be highly suppressed compared to the continuum photons.  Nonetheless,
the background rejection afforded by a monochromatic line signal easily compensates for
the reduction in the expected rates.  In this article, we will explore the null results
of searches for gamma ray lines by the Fermi LAT to constrain the properties of
dark matter interactions with the Standard Model.

If we are to assemble a coherent picture of what the inputs from direct detection, collider
searches, and indirect detection experiments imply about the nature of dark matter,
we need a framework which specifies how the WIMPs couple to various Standard Model
fields.  One popular way to proceed is to focus on a particular (usually pre-existing)
model containing a WIMP candidate, e.g. a supersymmetric theory with $R$-parity conservation
or a theory with Universal Extra Dimensions.  Such explorations are interesting, and probe
motivated visions for physics beyond the Standard Model.  However, they are often not generic.
Thus, rather than focus on a specific model, we choose to use the powerful machinery of
effective field theory (EFT) to capture features from a broad class of WMP models.
We specialize to the class of EFTs in which the relevant degrees of freedom are the dark
matter particle itself and the Standard Model -- a theory with a ``Maverick"
WIMP \cite{Beltran:2010ww}.

A Maverick WIMP couples to the Standard Model through high dimensional operators.  The
precise form of the interactions will depend on whether the dark matter is a fermion or a scalar.
These operators only accurately represent
 the physics in the regime where the EFT correctly captures all of
the relevant physics (at energies lower than the lightest degree of freedom after the dark matter
itself).  Provided this is true for all experiments of interest, the EFT provides a common language
which allows one to compare the constraints from different types of experiments.  This
approach has previously been fruitful in exploring the results of direct
detection \cite{Cao:2009uw,Beltran:2008xg}, collider
searches \cite{Cao:2009uw,Beltran:2010ww,Goodman:2010yf,Bai:2010hh,Shepherd:2009sa},
and continuum gamma ray production from WIMP annihilation
\cite{Cao:2009uw,Cirelli:2008pk}.
Here we extend the analysis to include indirect searches for gamma ray lines.

We begin in the next section by defining the EFTs and enumerating the set of
operators we consider.
We simplify our analysis by considering WIMPs which are singlets under SM gauge
transformations; the extension to other cases is straight-forward.
We proceed by finding the annihilation rate into two body final states including
photons (which typically involve a one loop Feynman diagram).   Given
our EFT assumptions, the allowed final states include $\gamma \gamma$, $\gamma Z$,
or $\gamma h^0$.  Since the SM Higgs has yet to be observed, and thus its mass is
unknown, we restrict our attention to the $\gamma \gamma$ and $\gamma Z$ cases.
We compare these rates to the bounds on gamma ray line searches
from 
Fermi, and find corresponding bounds on the dark
matter interactions. We also present the comparison to current bounds
coming from colliders and direct detection methods and finish with a discussion
of our results.

 \section{Effective  Theory  of WIMP interactions}
 \label{sec:couplings}

Following~\cite{Goodman:2010yf}, we assume that the WIMP $\chi$ is the only relevant
new particle in the energy range of interest, and thus the EFT contains the Standard Model
and $\chi$.  We restrict our attention to the cases where $\chi$ is a SM gauge singlet, but
allow it to be a real scalar, a complex scalar, a Majorana fermion, or a Dirac fermion.
We consider each of these spin cases separately, and we
ignore spin one or higher Lorentz representations.

Since the WIMP is neutral under the SM, the only way it can communicate with
the Standard Model fields is through higher dimensional operators\footnote{We assume that
the Higgs is heavy enough to be integrated out of the EFT, in which case it may be
responsible for UV-completing some of the operators which we do consider.} which factor into
a WIMP term multiplied by product of SM fields which itself forms a gauge singlet.  The two
factors are contracted together to form an over-all Lorentz invariant.
We are particularly interested in operators which couple the WIMP to quarks and gluons, since
these operators can produce signals at direct detection experiments.

The WIMP factors depend sensitively on the spin of $\chi$.  For a real scalar, only the factor
$\chi^2$ is non-vanishing; the $\chi \partial_\mu \chi$ construction can be integrated
by parts and moved entirely to the SM term.  For a complex scalar, $\chi^\dagger \chi$
and $\chi^\dagger \partial_\mu \chi$ are both independent.  Terms with more than one
derivative acting on $\chi$ are always subleading in the EFT.  For a fermion $\chi$,
the leading operators are of the form $\bar{\chi} \Gamma \chi$, where the matrices
$\Gamma$ are one of the set of $4 \times 4$ matrices, a complete basis for which is,
\bea
\left\{ 1, \gamma^5, \gamma^\mu, \gamma^\mu \gamma^5, \sigma^{\mu \nu} \right\}.
\label{eq:gammas}
\eea
Note that $\gamma^\mu$ and $\sigma^{\mu \nu}$ vanish for a Majorana $\chi$.
Terms with derivatives acting on a fermion WIMP are also subleading in the EFT.
In constructing the interactions for a Dirac fermion or complex scalar WIMP, we
restrict ourselves to those terms which leave a $U(1)_\chi$ symmetry unbroken,
as the lack of this symmetry would also allow for mass terms which would split the single
Dirac WIMP into two Majorana fermions or the complex scalar WIMP into
two real scalars.

The Standard Model factors consist of quark operators $\bar{q} \Gamma q$ (with the
color indices of the quarks contracted), where $\Gamma$ are matrices from the same
set as in Equation~(\ref{eq:gammas}), including:
\begin{itemize}
\item Quark scalar terms are constructed as
\bea
\sum_q ~m_q ~\bar{q} q ~~~~~ & {\rm or} & ~~~~~ \sum_q ~m_q~ \bar{q} \gamma_5 q~.
\eea
\item Quark vector terms are constructed:
\bea
\sum_q  ~\bar{q} \gamma_\mu q ~~~~~ & {\rm or} &
~~~~~ \sum_q ~\bar{q} \gamma_\mu \gamma_5 q~.
\eea
\item Quark tensor terms are constructed:
\bea
\sum_q  ~\bar{q} \sigma_{\mu \nu} q ~.
\eea
\end{itemize}
The constructions for scalar and vector bilinears may be further motivated
by minimal flavor violation \cite{Buras:2000dm}.  The normalization of the tensor
bilinear facilitates comparison with direct detection rates.
One could also write down analogous couplings to leptons.
In the case of the mass-suppressed operators this would have only a small effect
because leptons are light compared to quarks. In the cases where the mass of the
SM particle does not appear in the coupling the presence of a coupling to leptons with
identical strength would enhance the gamma ray line cross section by a factor of
$64/25$, but such operators have very
suppressed contributions to direct detection, and so we do not consider them here.
At one higher dimension, we also have dark matter couplings to
$G^a_{\mu\nu}G^a_{\alpha\beta}$, where $G^a_{\mu\nu} $
is the color field strength, and the Lorentz indices are contracted in all possible ways.
The gluonic operators are normalized by $\alpha_s$, which captures the dominant
renormalization group running and is motivated by the fact that such interactions are
most likely generated at the loop level
by integrating out a set of massive colored states.

\begin{table}[h]
 \hspace{0.033\textwidth}
 \begin{minipage}{0.4\textwidth}
  \centering
 \begin{tabular}{|c|c|c|}
\hline
   Name    & Operator & Coefficient  \\
\hline
D1 & $\bar{\chi}\chi\bar{q} q$ & $m_q/M_*^3$   \\
D2 & $\bar{\chi}\gamma^5\chi\bar{q} q$ & $im_q/M_*^3$    \\
D3 & $\bar{\chi}\chi\bar{q}\gamma^5 q$ & $im_q/M_*^3$    \\
D4 & $\bar{\chi}\gamma^5\chi\bar{q}\gamma^5 q$ & $m_q/M_*^3$   \\
D5 & $\bar{\chi}\gamma^{\mu}\chi\bar{q}\gamma_{\mu} q$ & $1/M_*^2$   \\
D6 & $\bar{\chi}\gamma^{\mu}\gamma^5\chi\bar{q}\gamma_{\mu} q$ & $1/M_*^2$    \\
D7 & $\bar{\chi}\gamma^{\mu}\chi\bar{q}\gamma_{\mu}\gamma^5 q$ & $1/M_*^2$   \\
D8 & $\bar{\chi}\gamma^{\mu}\gamma^5\chi\bar{q}\gamma_{\mu}\gamma^5 q$ & $1/M_*^2$   \\
D9 & $\bar{\chi}\sigma^{\mu\nu}\chi\bar{q}\sigma_{\mu\nu} q$ & $1/M_*^2$   \\
D10 & $\bar{\chi}\sigma_{\mu\nu}\gamma^5 \chi \bar{q}\sigma_{\mu\nu}q$ & $i/M_*^2$  \\
D11 & $\bar{\chi}\chi G_{\mu\nu}G^{\mu\nu}$ & $\alpha_s/4M_*^3$   \\
D12 & $\bar{\chi}\gamma^5\chi G_{\mu\nu}G^{\mu\nu}$ & $i\alpha_s/4M_*^3$   \\
D13 & $\bar{\chi}\chi G_{\mu\nu}\tilde{G}^{\mu\nu}$ & $i\alpha_s/4M_*^3$   \\
D14 & $\bar{\chi}\gamma^5\chi G_{\mu\nu}\tilde{G}^{\mu\nu}$  & $\alpha_s/4M_*^3$ \\
D15 & $\bar{\chi}\sigma^{\mu\nu}\chi F_{\mu\nu} $ & $M$   \\
D16 & $\bar{\chi}\sigma_{\mu\nu}\gamma^5 \chi F_{\mu \nu}$ & $D$  \\
\hline
M1 & $\bar{\chi}\chi\bar{q} q$ & $m_q/2M_*^3$   \\
M2 & $\bar{\chi}\gamma^5\chi\bar{q} q$ & $im_q/2M_*^3$    \\
\hline
\end{tabular}
  \label{tab:caption1}
 \end{minipage}
 \hspace{0.033\textwidth}
 \begin{minipage}{0.4\textwidth}
  \centering
\begin{tabular}{|c|c|c|}
\hline
   Name    & Operator & Coefficient   \\
\hline
M3 & $\bar{\chi}\chi\bar{q}\gamma^5 q$ & $im_q/2M_*^3$    \\
M4 & $\bar{\chi}\gamma^5\chi\bar{q}\gamma^5 q$ & $m_q/2M_*^3$   \\
M5 & $\bar{\chi}\gamma^{\mu}\gamma^5\chi\bar{q}\gamma_{\mu} q$ & $1/2M_*^2$    \\
M6 & $\bar{\chi}\gamma^{\mu}\gamma^5\chi\bar{q}\gamma_{\mu}\gamma^5 q$ & $1/2M_*^2$   \\
M7 & $\bar{\chi}\chi G_{\mu\nu}G^{\mu\nu}$ & $\alpha_s/8M_*^3$   \\
M8 & $\bar{\chi}\gamma^5\chi G_{\mu\nu}G^{\mu\nu}$ & $i\alpha_s/8M_*^3$   \\
M9 & $\bar{\chi}\chi G_{\mu\nu}\tilde{G}^{\mu\nu}$ & $i\alpha_s/8M_*^3$   \\
M10 & $\bar{\chi}\gamma^5\chi G_{\mu\nu}\tilde{G}^{\mu\nu}$  & $\alpha_s/8M_*^3$ \\ \hline
C1 & $\chi^\dagger\chi\bar{q}q$ & $m_q/M_*^2$    \\
C2 & $\chi^\dagger\chi\bar{q}\gamma^5 q$ & $im_q/M_*^2$   \\
C3 &  $\chi^\dagger\partial_\mu\chi\bar{q}\gamma^\mu q$ & $1/M_*^2$   \\
C4 &  $\chi^\dagger\partial_\mu\chi\bar{q}\gamma^\mu\gamma^5q$ & $1/M_*^2$    \\
C5 & $\chi^\dagger\chi G_{\mu\nu}G^{\mu\nu}$  & $\alpha_s/4M_*^2$   \\
C6 & $\chi^\dagger\chi G_{\mu\nu}\tilde{G}^{\mu\nu}$  & $i\alpha_s/4M_*^2$   \\ \hline
R1 & $\chi^2\bar{q}q$ & $m_q/2M_*^2$    \\
R2 & $\chi^2\bar{q}\gamma^5 q$ & $im_q/2M_*^2$   \\
R3 & $\chi^2 G_{\mu\nu}G^{\mu\nu}$  & $\alpha_s/8M_*^2$   \\
R4 & $\chi^2 G_{\mu\nu}\tilde{G}^{\mu\nu}$  & $i\alpha_s/8M_*^2$   \\
\hline
\end{tabular}
  \label{tab:caption2}
 \end{minipage}
   \caption{\textnormal{Operators coupling WIMPs to SM particles. The operator names
   beginning with D, M, C, R  apply to WIMPs that are Dirac fermions, Majorana fermions,
   complex scalars or real scalars respectively. }}
\end{table}

The coefficient of each operator is a dimensionful parameter, expressed in terms
of the appropriate power of the suppression scale $M_*$, which has dimensions
of [energy].  Each operator has its own $M_*$, but since we will consider one operator
at a time (which need not be the actual case but nonetheless captures the physics when one
operator dominates a given process) we refer to the set of parameters collectively as $M_*$.

In the case of a Dirac WIMP,
another possibility is a direct coupling of the dark matter to photons through either an
electric or magnetic  dipole interaction
\bea
D~\bar\chi \sigma_{\mu\nu}\gamma_5\chi ~F^{\mu\nu} ~~~~~ & {\rm or} &
~~~~~ M ~\bar\chi \sigma_{\mu\nu}\chi~ F^{\mu\nu}
\eea
where we refer to the coefficients of these two operators as $D$ and $M$, respectively.
As we shall see, these operators are probed particularly well by searches for gamma
ray lines due to the direct coupling of the dark matter to the photon
\cite{Sigurdson:2004zp,Bagnasco:1993st,Pospelov:2000bq,Gardner:2008yn,Cho:2010br}.
We note that there has been recent interest in dark matter with
dipole interactions, which have the potential to reconcile the DAMA signal
while remaining consistent with the null search results from CDMS and XENON
\cite{Masso:2009mu,Chang:2010en,Barger:2010gv,Banks:2010eh,Fitzpatrick:2010br}.

The complete list of operators that we consider is shown in Table I.  We adopt a naming
convention where the initial letter refers to the spin of $\chi$: D for Dirac fermion, M for Majorana,
C for complex scalar, and R for real scalar and the number specifies the particular
operator belonging to a given WIMP spin.  Within each family, the earlier numbers refer
to coupling to quark scalar bilinears (D1-4, M1-4, C1-2, and R1-2),
the middle numbers to quark vector bilinears (D5-8, M5-6, and C3-4)
and quark tensor bilinears (D9-10)
and the largest numbers to coupling to gluons (D11-14, M7-10, C5-6, and R3-4).  The
WIMP electric and magnetic dipole moment operators are labelled D15 and D16.

\section{Gamma Ray Line Search Constraints}

\begin{figure}[t]
\includegraphics[width=6.0cm]{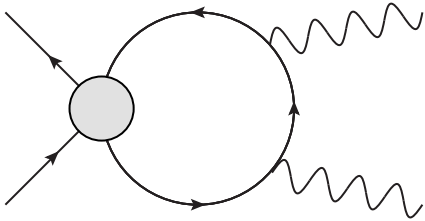}
\caption{\label{fig:diagram}
Representative Feynman diagram for the loop level annihilation of two DM
particles $\chi$ to a photon and a second vector boson, either another photon or a $Z$ boson,
through an operator coupling the DM to SM quarks (represented as the shaded circle).
}
\end{figure}

We compute the rate for the processes $\chi\chi\to\gamma\gamma$ and
$\chi\chi\to\gamma Z$ for
each of the operators considered above.  Generally, stronger bounds arise from the
$\gamma \gamma$ process because it produces two photons per annihilation (compensating
for the $Z$ coupling to quarks being typically a little stronger than the photon).  Consequently,
we consider the $\gamma Z$ final state only in the case where annihilation into
$\gamma \gamma$ vanishes.
For the cases with a Dirac fermion or complex scalar,
we assume that the dark matter in our galactic halo is composed of equal numbers of
particles and anti-particles.  It should be borne in mind that one could evade the constraints
from any annihilation process if the interactions preserve the
$U(1)_\chi$ symmetry and the galactic halo is made entirely of WIMPs or anti-WIMPs.

For the operators D15 and D16 mediating a direct interaction between the WIMPs and the
photon, this process occurs at tree level.  Generally, the quark operators mediate
annihilations into $\gamma \gamma$ or $\gamma Z$
at the one loop level as shown in Figure~\ref{fig:diagram}.  For the
operators of the form $\bar{\chi}\Gamma^\mu\chi\bar{q}\gamma_\mu q$, a final state
containing two photons is forbidden by the Landau-Yang theorem \cite{Yang:1950rg}.
For these operators, we rely on $\chi \chi \to \gamma Z$ to determine the implications of
searches for gamma ray lines.
For operators coupling the WIMPs directly to gluons and for the tensor operators D9 and D10,
the leading contribution to $\gamma \gamma$ and $\gamma Z$ final states occurs at
two loops, and as a result the rate
is expected to be small enough that these operators are much better probed by
collider searches \cite{Goodman:2010yf}.  Thus, we do not consider
indirect bounds on these operators.

The one loop amplitudes were generated using the FeynArts package\,\cite{Hahn:2000kx}
from which cross sections were derived using FormCalc\,\cite{Hahn:1998yk}.
As a cross check, the results for the operators D2 and D4 were compared
at the analytic level of the scalar loop integrals
(after the Passarino-Veltman reduction \cite{Passarino:1978jh}
was performed by hand), and agreement found with the FeynArts results.  Despite the fact
that our EFT is nonrenormalizable, all of the one loop amplitudes for
annihilation into $\gamma \gamma$ and
$\gamma Z$ are finite, insensitive to the detailed UV theory.

Armed with the cross sections for WIMP annihilation into $\gamma \gamma$ or $\gamma Z$,
we compare with the bounds from the 
Fermi LAT null searches for gamma ray lines.
The Fermi LAT
has released the results of searches from 30 GeV to 200 GeV \cite{Abdo:2010nc}, and
is eventually expected to reach lower masses as well.

In terms of the annihilation cross section and WIMP mass, the flux produced in
monochromatic lines is given by
\bea
\Phi_S &=& \frac{1}{4 \pi} \frac{r_{\odot} \rho^2_{\odot}}{m_\chi^2}~
\langle\sigma (\chi \chi \rightarrow \gamma \gamma) v\rangle
~ \bar{J} \times \left( \Delta \Omega \right)
\eea
where $\Delta \Omega$ is the size of the angular acceptance and
$\bar{J}$ encodes the integral of the dark matter density $\rho$ squared along the line of sight,
\bea
\bar{J} & \equiv & \frac{1}{\Delta \Omega}
\int_{\Delta \Omega} d\Omega \int_{\rm LOS}
\frac{dl}{r_\odot}~ \frac{\rho^2 (l,\Omega) }{\rho^2_\odot}
\eea
with $r_\odot = 8.5$~kpc the distance from the sun to the galactic center and
$\rho_\odot$ the local density of dark matter.
For
annihilations into $\gamma Z$, the final state photon multiplicity is reduced by a factor of
two, and the energy of the line is mapped onto the WIMP mass via Eq.~(\ref{eq:Eline}) with
$M_X = M_Z$.

The quantity $\bar{J}$
is typically derived from $N$-body simulations simulating the formation and evolution
of Milky Way-like galaxies.  However, the simulations cannot resolve scales of order the
galactic center and do not include effects from baryonic matter which may strongly
affect the dark matter profile close to the core.  As a result, it represents the largest uncertainty
in mapping the experimental results onto cross sections (and from there onto the coefficients
of operators).  Typical choices of the dark matter density thought to provide realistic
pictures of the milky way halo are the Navarro-Frenk-White NFW \cite{Navarro:1995iw} or
``Einasto" \cite{Graham:2005xx} profiles.
In order to derive the most conservative bounds, we choose the Fermi limits
corresponding to
an isothermal
profile \cite{Bahcall:1980fb}
which results in a modest $\bar{J}$
and correspondingly weaker limits (about a factor of two compared to NFW for the
Fermi line search region of interest \cite{Abdo:2010nc})
on the annihilation cross section.

\begin{figure}[t]
\includegraphics[width=12.0cm]{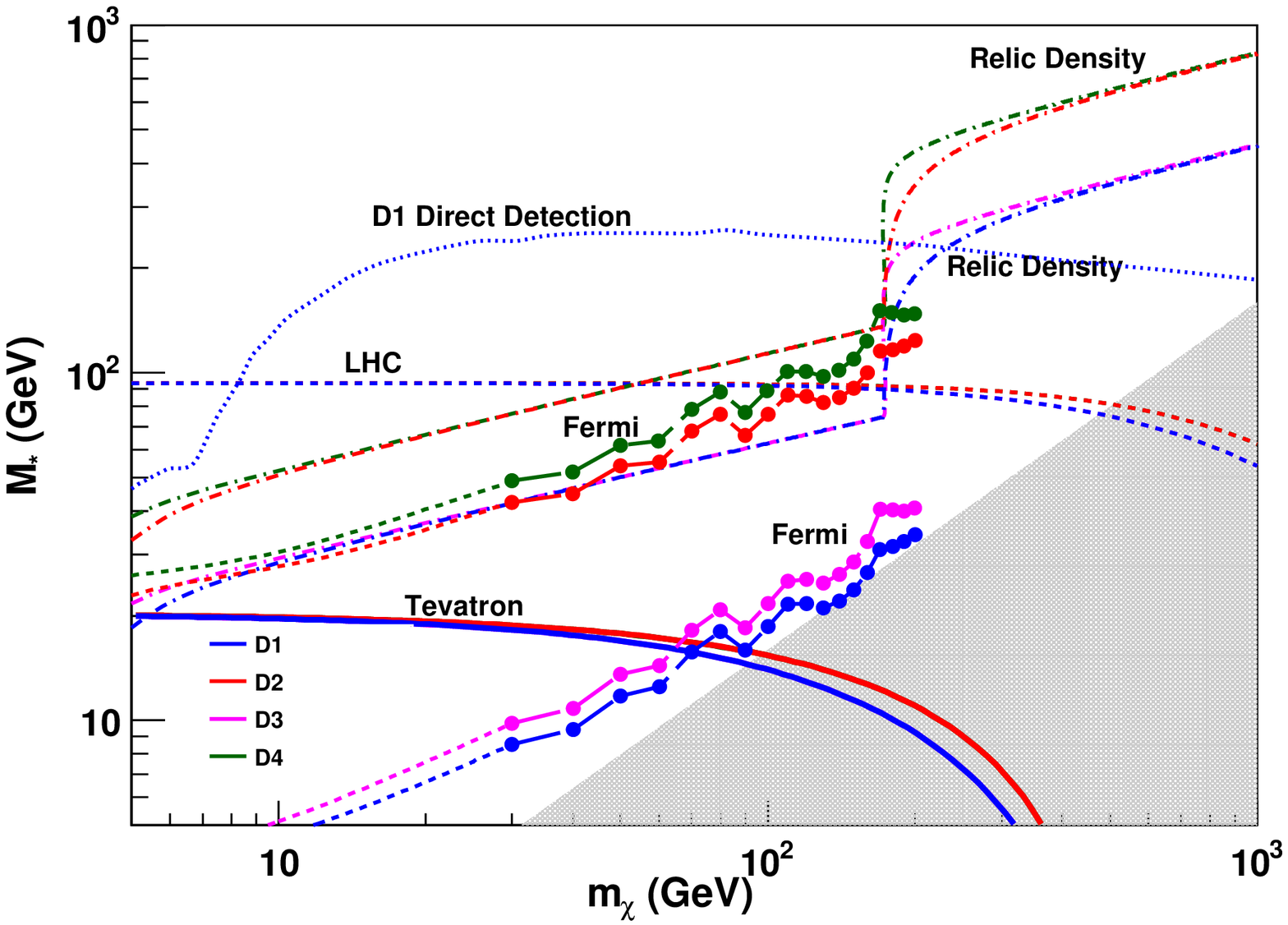}
\caption{\label{fig:D1-4}
The lower limits from current Fermi line searches\,\cite{Abdo:2010nc} (long dashed lines with
data points shown) and an estimate of the reach of future searches at lower energies (short
dashed curve) on the suppression scale of new physics $M_*$ leading to interactions with the SM
for the Dirac WIMP operators D1-D4.
Note that the constraints on D1 and D3 are significantly
weaker than the others, because these operators lead to cross sections which are
velocity-suppressed.
For comparison we also show the current bounds from
Tevatron and future reach of LHC\,\cite{Goodman:2010yf}
(solid and short dashed curves, respectively), as
well as the value of $M_*$ leading to the correct thermal relic abundance in the absence of
other interactions (dash-dotted curves).
}
\end{figure}

\begin{figure}[t]
\includegraphics[width=12.0cm]{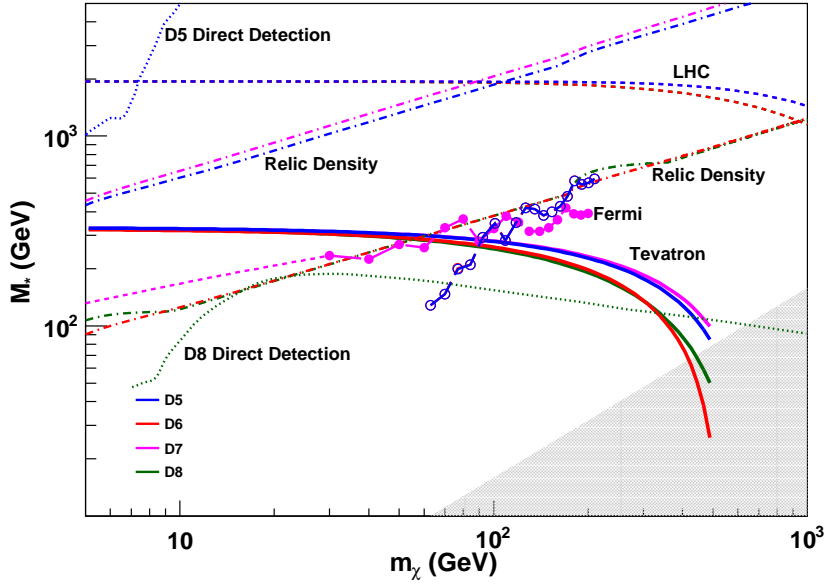}
\caption{\label{fig:D5-8}
Same as figure\,\ref{fig:D1-4}, but for the Dirac WIMP operators D5-D8. Constraints for D5
and D6 arise from the $\gamma Z$ annihilation process, and thus there are
no effective line search bounds
for masses $m_\chi\lesssim50$ GeV.
}
\end{figure}

\begin{figure}[t]
\includegraphics[width=12.0cm]{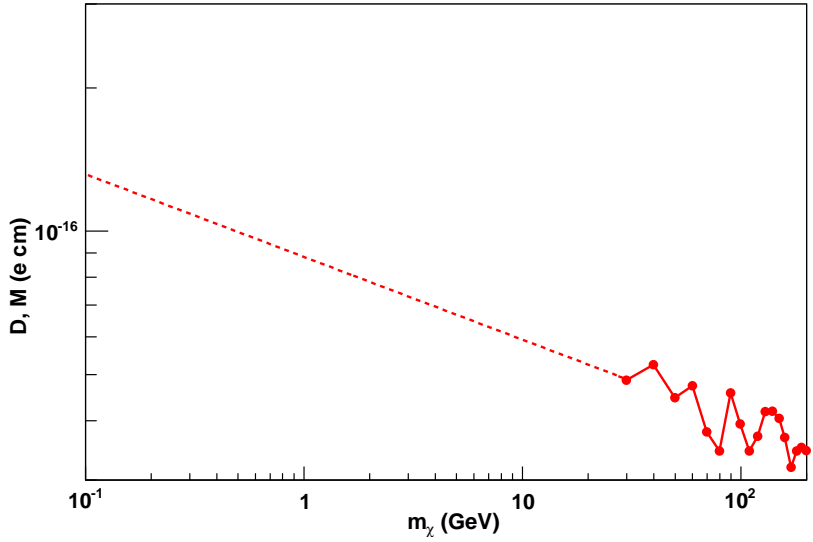}
\caption{\label{fig:dipole}
Experimental upper bounds on dipole moments of dark matter (operators D15 and D16)
derived from the spectral line searches by 
FERMI\,\cite{Abdo:2010nc} (solid lines). The dashed curve is our estimate of
the reach of a future Fermi line search covering the illustrated lower energy
range.
}
\end{figure}

\begin{figure}[t]
\includegraphics[width=12.0cm]{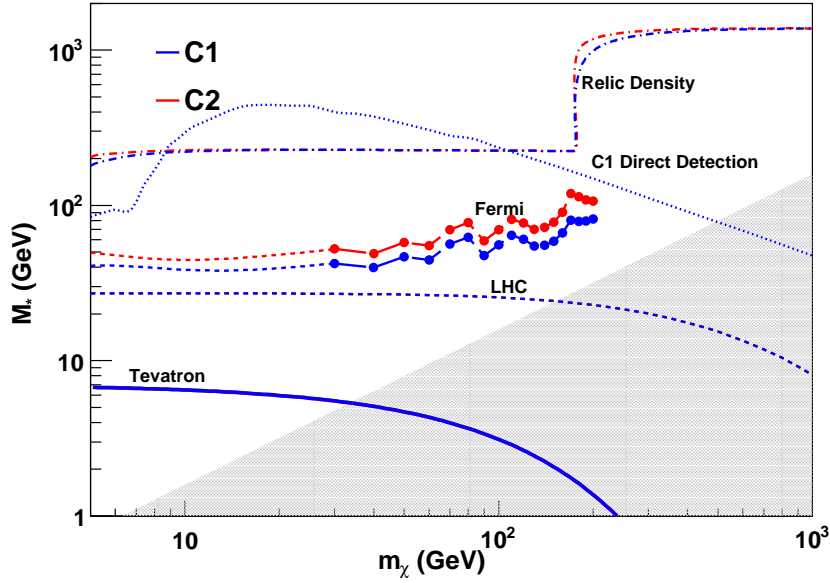}
\caption{\label{fig:C1-2}
Same as figure\,\ref{fig:D1-4}, but with constraints shown for the complex scalar WIMP
interacting with the SM through operators C1 and C2.
}
\end{figure}

\begin{figure}[t]
\includegraphics[width=12.0cm]{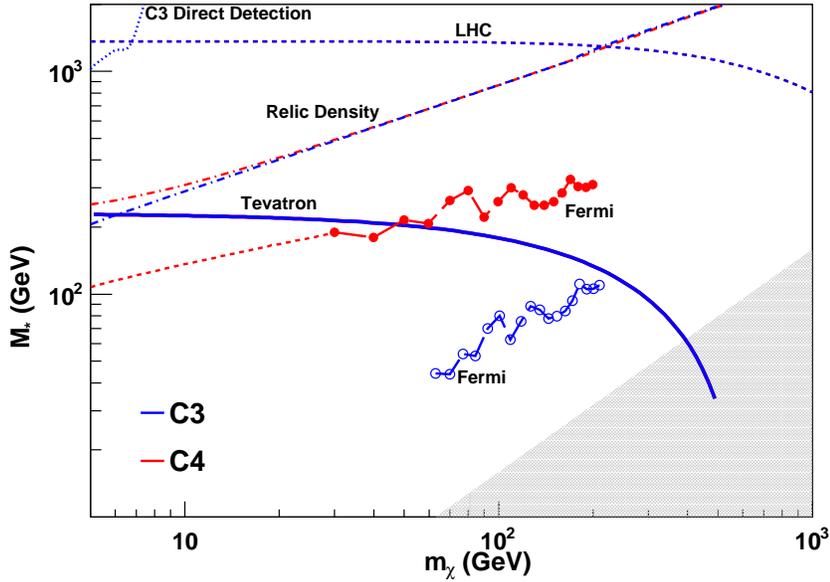}
\caption{\label{fig:C3-4}
Same as figure\,\ref{fig:D1-4}, but with constraints shown for the complex scalar WIMP
interacting through operators C3 and C4. Operator C3 is only constrained by the
$\gamma Z$ annihilation process, and thus there are no effective line search bounds for masses
$m_\chi\lesssim50$ GeV.
}
\end{figure}

The null results of searches for gamma ray lines, together with our choice of the isothermal
dark matter distribution, puts a $95\%$ CL bound on on the
cross section $\sigma (\chi \chi \rightarrow \gamma \gamma)$
(or in some cases $\sigma (\chi \chi \rightarrow \gamma Z)$).  These are mapped into a
bound on the parameter $M_*$ which appears in the coefficients of each operator.
These bounds\footnote{The bounds on Majorana and
real scalar WIMPs are simply related to the corresponding bounds from Dirac and
complex scalar WIMPs respectively, once one notes that the bounds on the identical
particle cross sections are a factor of two better than the corresponding bounds on
$\chi \bar{\chi}$ (particle-antiparticle) annihilation.}
are presented in Figures~\ref{fig:D1-4}-\ref{fig:C3-4}, along with
the region where each operator provides the correct thermal relic
density and the previous $95\%$ CL
bounds from Tevatron data (as well as the future prospects
for a $5\sigma$ detection at the LHC)
\cite{Goodman:2010yf}.

We find that for the cases of operators
D1 and D3, the line searches from Fermi are actually stronger than collider bounds
for WIMP masses greater than about 50 GeV and for D2 and D4, are stronger
than the Tevatron bounds for all WIMP masses covered by the Fermi line search
and stronger than the LHC prospects for masses greater than about 100 GeV.
For operators D5 and D7, Fermi bounds are stronger for masses greater than about
90 GeV.  For operators C1, C2, and C4, the line searches are
stronger than the Tevatron bounds for larger WIMP masses,
with C1 and C2 much more strongly probed than is even possible at the LHC.
Overall, our results show a rich potential for synergy between different types of
experimental probes of dark matter.

For the cases where the dark matter couples directly to photons
(operators D15 and D16),
the existing bounds from
 Fermi  are shown in Fig.~\ref{fig:dipole}
as the
solid red line beyond masses of 30 GeV.
While the current Fermi limits only go down to lines of energy 30 GeV, the Fermi LAT is
also potentially sensitive to much smaller energies.  In the absence of a published study,
we extrapolate the existing bounds of \cite{Abdo:2010nc}
to lower masses.  While these extrapolations are
hopefully semi-realistic at masses not much below the current lower limit of 30 GeV, we caution
that for very low masses they may be unrealistic
and/or unconservative.  Obviously, the most ideal situation would
be to have updated limits by Fermi itself in the future.  To estimate the background,
we assume that the background is described by a power law below 30 GeV.  The
data \cite{Abdo:2010nc} between 30 and 55 GeV are described by a power law,
\beq
\frac{dB}{dE}=1.22\times10^6 {\rm~GeV^{1.39}}\times E^{-2.39}
\label{eq:Bextrap}
\eeq
where $dB / dE$ is the differential number of events (corresponding to an 11 month
exposure) in energy.  We treat the energy resolution of
the LAT as ${\Delta E / E} = 10\%$, constant in energy.  Choosing a bin
size corresponding to twice the experimental resolution, we find the expected number of
background events in the bin by integrating the background function,
Eq.~(\ref{eq:Bextrap}).
We place $95\%$ confidence level expected limits on a hypothesized signal line
producing $S$ events by requiring $S / \sqrt{B} < 1.96$.
As a check of the consistency of our parameterization, we derive the flux bounds
in \cite{Abdo:2010nc} and find they are consistent with the actual
experimental results to within better than $5\%$ over the entire range 30-200~GeV.
The resulting expected bounds on $M_*$, $D$, and $M$ from the extrapolated
Fermi line search are plotted in
Figures~\ref{fig:D1-4}-\ref{fig:C3-4} as dashed lines extending below 30 GeV.

\section{Discussion and Conclusions}

\begin{figure}[t]
\includegraphics[width=17.0cm]{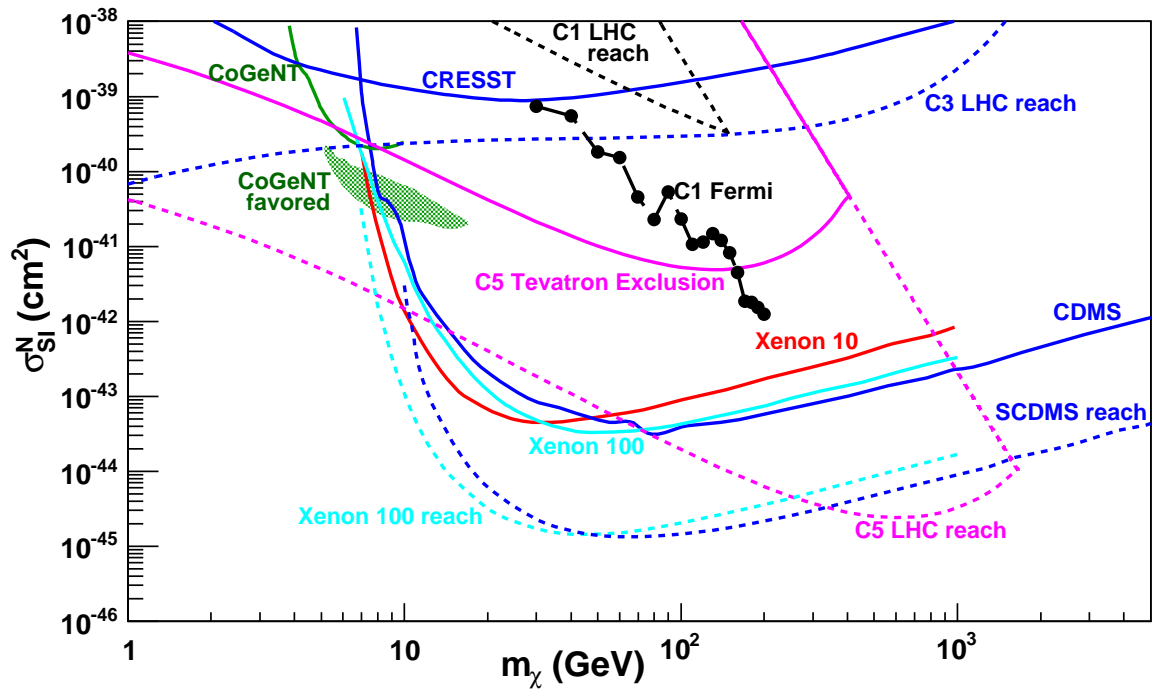}
\caption{\label{fig:si-plot}
Graph of the parameter space for spin-independent direct detection of WIMPs, in terms of the
WIMP mass and cross section for scattering on a nucleon, including current bounds from
direct detection, Tevatron data, and the Fermi line search (solid lines, as labelled), and projected
lines for direct detection experiments and the LHC (dashed lines, as labelled). The bounds on complex scalars are shown. The bounds for the operator C3 are above the axes displayed here. Fermi bounds on real scalars are obtained by improving the corresponding bounds for the complex scalars by a factor of 2. Fermi bounds on fermion operators are weak and outside the scale of this plot.
}
\end{figure}

\begin{figure}[t]
\includegraphics[width=17.0cm]{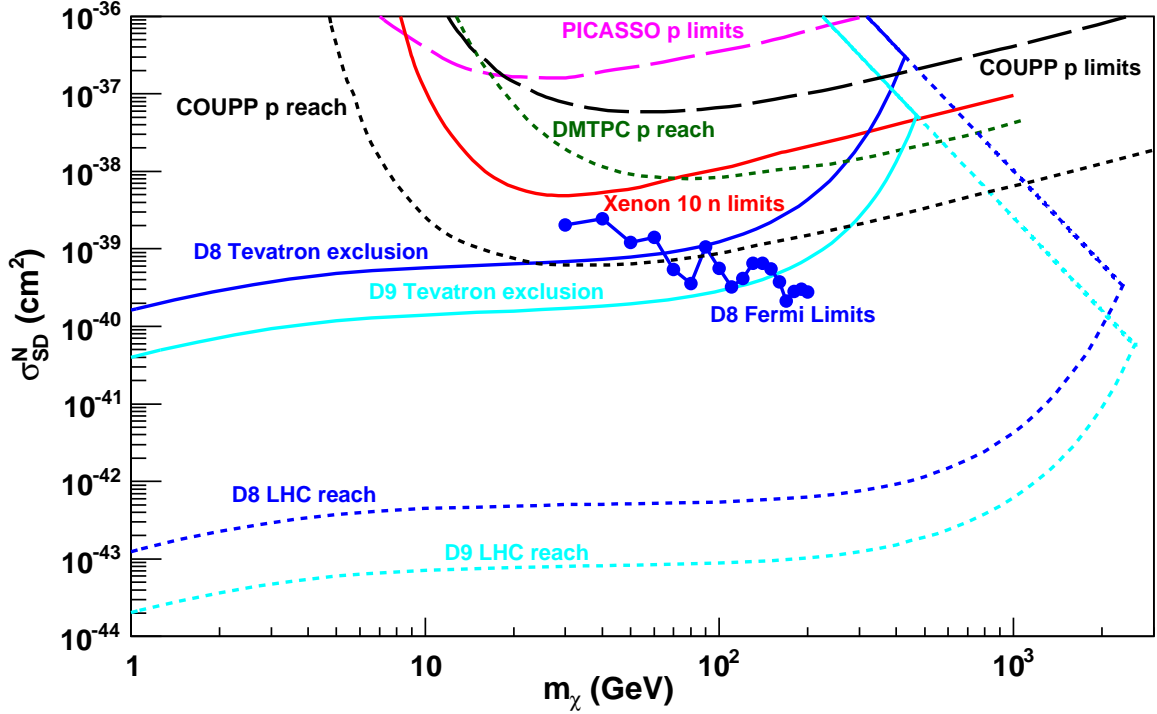}
\caption{\label{fig:sd-plot}
Graph of the parameter space for spin-dependent direct detection of WIMPs, in terms of the
WIMP mass and cross section for scattering on a proton or neutron, including current bounds from
direct detection, Tevatron data, and the Fermi line search (solid lines, as labelled), and projected
lines for direct detection experiments and the LHC (dashed lines, as labelled). Fermi bounds on Majorana fermion with M5 interaction are obtained by improving the D8 bounds for the Dirac fermion by a factor of 2.
}
\end{figure}

\begin{figure}[t]
\includegraphics[width=17.0cm]{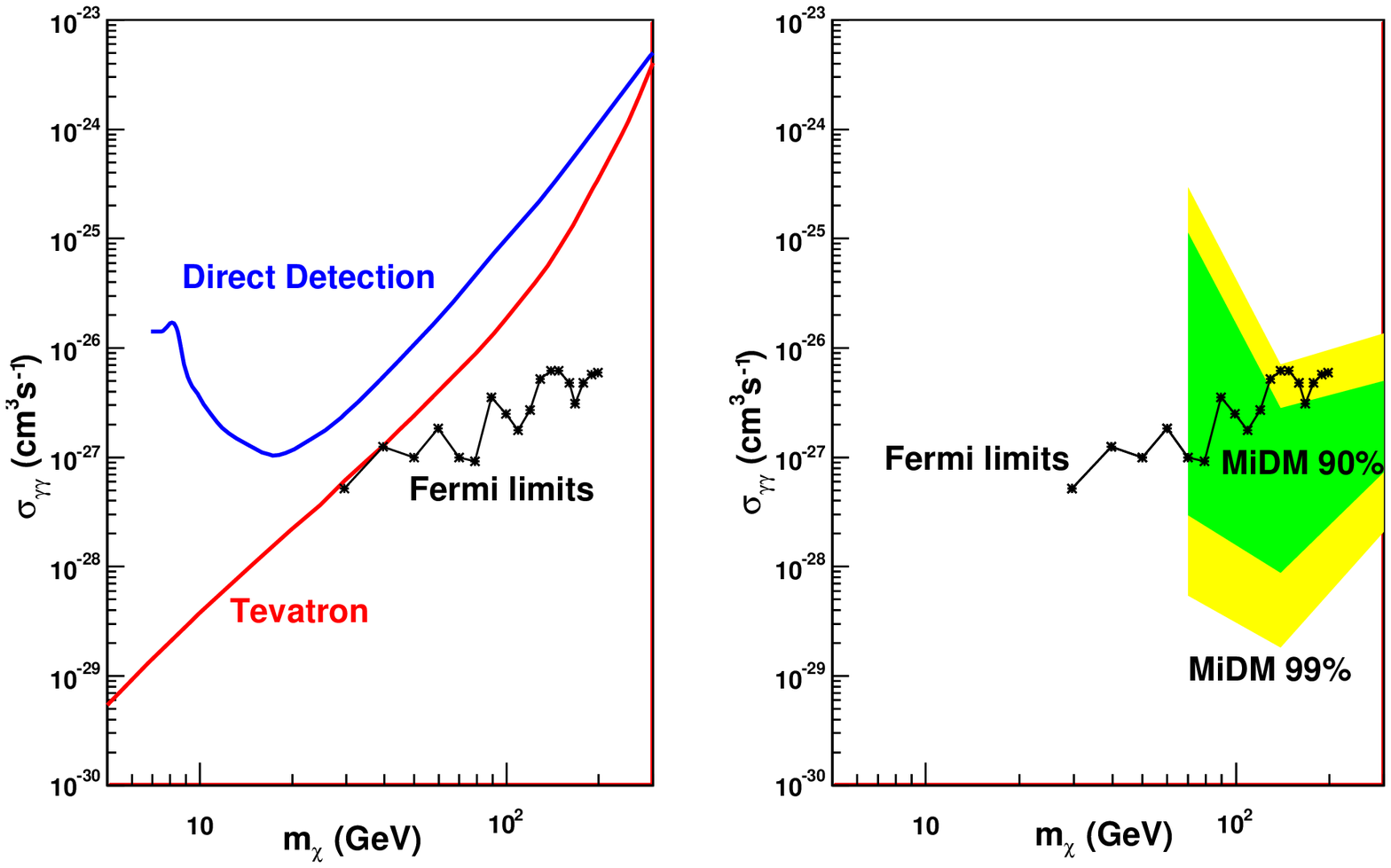}
\caption{\label{fig:id-plot}
On the left, a graph of the parameter space for indirect detection of gamma ray lines
for a Majorana WIMP interacting primarily through operator M6, in terms of the
WIMP mass and cross section annihilation into $\gamma \gamma$.  Shown are
the Fermi line search bounds, bounds from
Tevatron data, and the direct detection experiments
(solid lines, as labelled).  On the right, a comparison between the Fermi line search bounds and
the  MiDM \protect{\cite{Chang:2010en} model parameter space. The shaded regions correspond to the $90\%$ ($99\%$)
fits of the MiDM model to explain the DAMA signal }.
}
\end{figure}

The power of an effective theory description of WIMP interactions
is that it also allows one to make predictions for
any observable in the domain of validity of the effective theory.
In Figure~\ref{fig:si-plot}, we show the parameter space for a complex
scalar WIMP scattering through a spin-independent interaction with a nucleon
along with some of the existing bounds and prospects for direct detection experiments
\cite{Aalseth:2010vx,Aprile:2010um,Ahmed:2009zw}.  The limits
from gamma ray lines are worse than the current limits from Xenon 10 and CDMS, but are
of the same order for WIMP masses around 200 GeV.  If the LAT were to observe
a gamma ray line signal in the near future, this would (for example) exclude these
combinations of WIMP masses and couplings as being responsible.
In Figure~\ref{fig:sd-plot}, we show the parameter space for a Dirac WIMP scattering
through a spin-dependent interaction along with some of the relevant direct
detection experiments \cite{Archambault:2009sm}.  We see that, as was also true for colliders,
the line searches are stronger for some operators than near future prospects for
spin-dependent scattering.  Again, in the presence of a signal this would favor some WIMP
spins and operators over others.

In Figure~\ref{fig:id-plot} we present the parameter space for a Majorana WIMP annihilating into
two photons through the operator M6, in the parameter space of $m_\chi$ and
$\sigma (\chi \chi \rightarrow \gamma \gamma)$.  Together with the Fermi line search data,
we have shown the best limits from direct detection experiments \cite{Archambault:2009sm}
and the Tevatron limits \cite{Goodman:2010yf}.  For M6, the Fermi line search is currently
the best available limits for the masses to which they apply, with the Tevatron limits
comparable and taking over at low WIMP masses.  Direct detection limits become the
most constraining for WIMP masses above $\gsim 300$~GeV.

Also shown in Figure~\ref{fig:id-plot} is the parameter space for which
``Magnetic inelastic Dark Matter" (MiDM) \cite{Chang:2010en}, in its electromagnetic
(as opposed to dark photon) incarnation explains the DAMA signal.  In this
model the WIMP $\chi$ is a Majorana fermion which scatters
inelastically into a slightly heavier state $\chi^*$ through a flavor-changing
magnetic dipole moment,
\bea
M ~\bar{\chi} \sigma^{\mu \nu} \chi^* ~F_{\mu \nu} + H.c.
\eea
This operator also induces a line signal when two $\chi$ (or two $\chi^*$) annihilate
into $\gamma \gamma$ by exchanging a $\chi^*$($\chi$).
The regions correspond to $90\%$ and $99\%$ CL consistent parameter spaces in
$m_\chi$ and $M$ (as found in \cite{Chang:2010en}), mapped into the
parameter space of indirect detection under the assumption that the
dark matter halo is composed entirely of $\chi$ or $\chi^*$.  The Fermi line search is already
providing interesting constraints on the model parameter space.

Effective theories have the potential to be powerful descriptions of WIMP interactions.  In this
work, we have explored how the bounds from gamma ray lines map into constraints on
the nature of these interactions, and seen how these constraints are an interesting,
complementary picture to those provided by direct detection and colliders.  All together,
they provide a multi-faceted probe of the nature of dark matter.

\section*{Acknowledgements}
The authors are grateful for conversations with E. Do Couto E Silva and S.~Murgia.
T. Tait acknowledges the hospitality of the SLAC theory group, where part of this work was
completed.  This research is supported in part by NSF Grants No. PHY-0653656
and PHY--0709742.

\end{document}